\documentclass[onecolumn]{article}
\usepackage{amsmath, amsthm, amscd, amsfonts, amssymb, graphicx, hhline}
\usepackage{dcolumn}
\usepackage{bm}
\usepackage{multirow}
\usepackage[]{color}
\usepackage[papersize={210.0mm,297.0mm},dvips]{geometry}
\providecommand{\keywords}[1]
{
  \begin{quote}
  \small Keywords: #1
  \end{quote}
}
\providecommand{\pacs}[1]
{
  \begin{quote}
  \small PACS numbers: #1
  \end{quote}
}

\begin{document}
\title{Role of the equivalence principle in gauge and axial symmetries of Yukawa coupling, and the strong CP problem}
\author{Konstantin V. Grigorishin\footnote{\small konst.phys@gmail.com}\\ \small \textit{Bogolyubov Institute for Theoretical Physics of the National Academy}\\ \small \textit{of Sciences of Ukraine, 14-b Metrolohichna str. Kyiv, 03143, Ukraine}}
\date{}
\maketitle
\begin{abstract}
It is demonstrated the fundamental role of the equivalence principle in gravity for the Yukawa coupling between scalar and fermion fields. The Kibble-Zurek mechanism for formation of topological defects as vortexes and monopoles breaks down in system with a global gauge symmetry only. At the same time, the different vacuums can occur, which are separated be domain walls. The equivalence principle makes the strong violation of CP invariance impossible. Thus the axion hypothesis becomes redundant.
\end{abstract}

\normalsize
\section{Introduction}\label{intro}

There is a deep analogy between condensed matter physics (CMP) and particle physics: in both sections, the spontaneous symmetry breaking occurs accompanied by the Higgs effect and appearence of Goldstone bosons. For example, ferromagnetism is consequence of spontaneous breaking of rotational symmetry of initial Hamiltonian, and magnons play role of the Goldstone bosons. This analogy is most clearly manifested in superconductors \cite{sad,ryder}. Ginzburg-Landau (GL) theory is, in fact, an Abelian Higgs model, however stationary: the superconducting order parameter $\langle aa\rangle\propto|\varphi|e^{\mathrm{i}\theta},\langle a^{+}a^{+}\rangle\propto|\varphi|e^{-\mathrm{i}\theta}$ plays role of the complex scalar field, the density of free energy plays role of Lagrangian, the phase $\theta(\mathbf{r})$ play role of the Goldstone boson, which is absorbed by magnetic field $\mathbf{A}$, and the equation of field has a form $\nabla^{2}\mathbf{A}=1/\lambda^{2}\mathbf{A}$, where the reciprocal of the magnetic penetration depth $\lambda$ plays role of mass of photons, i.e the Higgs effect takes place. In Ref.\cite{grig1} the dynamical Lorentz-covariant (where speed of order of Fermi speed $v=v_{F}/\sqrt{3}$ plays role of the light speed) generalization of GL theory is formulated, where there is analog of Higgs boson - oscillations of the module of order parameter $|\varphi|$, and the Goldstone oscillations $\theta(t,\mathbf{r})$ are absorbed by electro-magnetic field $A_{\mu}$ so that a photon obtains mass. However, the boundary conditions on the surface of the superconductor lead to complete screening of the electrostatic field, as in the case of an ordinary metal. In Ref.\cite{grig2,grig3} these results have been generalised for two- and three-band superconductors with several order parameters $|\varphi_{i}|e^{-\mathrm{i}\theta_{i}},i=1,2,3$ with interband (Josephson) coupling $\epsilon(\varphi_{i}^{+}\varphi_{j}+\varphi_{i}\varphi_{j}^{+})=2\epsilon|\varphi_{i}||\varphi_{j}|\cos(\theta_{i}-\theta_{j})$, where it has been demonstrated, that the Goldstone modes from each band transform into normal oscillations for all bands: common mode oscillations $\theta_{i}$ with acoustic spectrum, which are absorbed by the gauge field, and anti-phase oscillations $\theta_{i}-\theta_{j}$ with an energy gap in spectrum (mass) determined with the interband coupling $m_{L}\sim\sqrt{|\epsilon|}$, which can be associated with the Leggett modes. In Ref.\cite{grig} it has been proposed an extension of the Glashow-Weinberg-Salam model of electro-weak interaction using analogy with three-band superconductors with interband Josephson couplings between strongly asymmetrical condensates of scalar fields. In this model there are two sterile ultra-light Leggett bosons which are possible particles of "dark matter".

At the same time this analogy is not complect. In the particle physics there is Yukawa coupling between scalar field and fermions, which is term of the form $\chi(\overline{\psi}_{L}\varphi\psi_{R}+\overline{\psi}_{R}\varphi^{+}\psi_{L})$, where $\psi_{L}$ and $\psi_{R}$ are left-handed and right-handed spinor accordingly, $\chi$ is the dimensionless coupling constant. In CMP such an interaction is absent. The Yukawa coupling causes Dirac mass of the fermion $m_{D}=\chi|\varphi|$. However, there is another significant effect, which is the so-called strong CP problem: Yukawa coupling in a CP non-invariant form $m_{D}(\overline{\psi}_{L}e^{2i\beta}\psi_{R}+\overline{\psi}_{R}e^{-2i\beta}\psi_{L})$ cannot be reduce to $m_{D}(\overline{\psi}_{L}\psi_{R}+\overline{\psi}_{R}\psi_{L})$ through axial transformation $\psi_{L}\rightarrow e^{i\beta}\psi_{L},\psi_{R}\rightarrow e^{-i\beta}\psi_{R}$ because the Lagrangian is non-invariant with respect to such transformation due to quantum effects. Then, a neutron must have an electric dipole moment $\mathbf{d}$ since this violates CP symmetry (but remains CPT symmetry). Measurements give $d=(0.0\pm1.1_{\mathrm{stat}}\pm0.2_{\mathrm{sys}})\cdot 10^{-26}\mathrm{e}\cdot\mathrm{cm}$ \cite{abel}, that is it is zero. This fact requires explanation (we can choice $\beta=0$ but this is an unjustified artificial selection). It is proposed the Peccei–Quinn mechanism, where the axial symmetry is made local but is spontaneously broken, giving rise to a pseudo-Goldstone boson $\theta$ called an axion, so that the axion condensate $\langle\theta\rangle$ eliminates the phase $\beta+\langle\theta\rangle/(2N_{f})=0$, where $N_{f}$ is number of types of quarks. It is noteworthy that axions are considered viable candidates for Dark Matter. However, the axion hypothesis makes more problems than it solves: the existence of axions would lead to additional energy emission from stars and white dwarfs, which contradicts observational data, in addition, axions directly have not yet been detected in the collider experiments or in cosmic rays. There are other models (non Peccei–Quinn mechanism) solving the strong CP problem. However all these models require some kind of extensions of the Standard Model with all the problems that entails.

A significant property of CMP is the presence of topological objects (defects). In a number of cases, such objects contain at least one point where the order parameter is zero, despite the fact that the system's symmetry is spontaneously broken throughout space. For example, domain wall in ferromagnetic - topological defect, where directly on the wall the magnetization (order parameter) is zero $\mathbf{M}=0$, i.e $|\mathbf{M}|=0$ and, accordingly, the direction of vector $\mathbf{M}$  is undefined. In superconductors and superfluid helium the quantum vortexes can exist. Along central axis of each vortex we have region of thickness of the coherence length $\xi$ where $\varphi=0$, i.e $|\varphi|=0$ and the phase $\theta$ is undefined. It is not difficult to see that such $2D$ (walls) and $1D$ (vortexes) defects make the system highly inhomogeneous and anisotropic.


As in CMP, in quantum field theory there are solutions of field equations which are topological defects. The most notable of these are: monopoles, cosmic strings and domain walls (in the case of discrete symmetry) \cite{vilen,rubakov}. The strings and walls are $1D$ and $2D$ objects accordingly, therefore their presence makes space essentially inhomogeneous and anisotropic, that contradicts observational data. In addition, the presence of monopoles could not be detected despite numerous attempts. This problem is currently being addressed using the inflation hypothesis \cite{linde,dolgov}: one domain exponentially quickly inflates $a=a_{0}e^{Ht}$ to enormous sizes $a\ggg a_{0}$ (here $H\approx \mathrm{const}$ is Hubble parameter), so that an observer in the domain will observe a homogeneous and isotropic space, in addition, the concentration of monopoles will be negligibly small. However the inflation generates gravitational waves \cite{dolgov}, which play role the relic gravitational waves now, which have never been discovered despite the efforts made. This may indicate that there are non-inflationary mechanisms to ensure the absence of the above-mentioned topological defects. In connection with the above, the following circumstance should be noted: in the particle physics the gravitational field is considered exclusively as a background against which gauge and other interactions take place due to the fact that gravitational interaction is incomparably weaker than other types of interactions. However, the gravity is, first and foremost, symmetry: in the infinitely small vicinity of each point, space can always be made flat by an appropriate choice of reference frame (the equivalence principle). As will be demonstrated in the paper, the Yukawa coupling together with the equivalence principle play a fundamental role in ensuring of the strong CP invariance (leaving only the possibility for its weak violation) and in formation of topological defects.

\section{Electromagnetic $U(1)$ symmetry}\label{U1}

Let us consider an electromagnetic group $U(1)$ as a so-called toy model. We have complex scalar field, which are equivalent to two real scalar fields each: modulus $\left|\varphi(x)\right|$ and phase $\theta(x)$ (the modulus-phase representation):
\begin{equation}\label{1.1}
    \varphi(x)=\left|\varphi(x)\right|e^{\mathrm{i}\theta(x)}.
\end{equation}
Here $x\equiv(t,\textbf{r})$, and we will use the system of units, where $c=\hbar=1$. Let us consider the system with Lagrangian:
\begin{eqnarray}\label{1.2}
    \mathcal{L}&=&L_{\varphi}+L_{\psi}+L_{A}\equiv(\partial_{\mu}+\mathrm{i}eA_{\mu})\varphi(\partial^{\mu}-\mathrm{i}eA^{\mu})\varphi^{+}
    -a\left|\varphi\right|^{2}-\frac{b}{2}\left|\varphi\right|^{4}\nonumber\\
    &+&\mathrm{i}\overline{\psi}_{L}\gamma^{\mu}(\partial_{\mu}+\mathrm{i}eA_{\mu})\psi_{L}
    +\mathrm{i}\overline{\psi}_{R}\gamma^{\mu}(\partial_{\mu}+\mathrm{i}eA_{\mu})\psi_{R}-\frac{1}{16\pi}F_{\mu\nu}F^{^{\mu\nu}},
\end{eqnarray}
where $\partial_{\mu}\equiv\frac{\partial}{\partial x^{\mu}}\equiv\left(\frac{\partial}{\partial t},\nabla\right),
\quad\partial^{\mu}\equiv\frac{\partial}{\partial x_{\mu}}\equiv\left(\frac{\partial}{\partial t},-\nabla\right)$ are covariant and contravariant differential operators accordingly. The coefficient $a<0$ and the coefficient $b>0$, hence the field condensate $\varphi_{0}=\sqrt{\frac{-a}{b}}$ is present. $\gamma^{\mu}$ are Dirac matrices, 
$\overline{\psi}=\psi^{+}\gamma^{0}$ is the Dirac conjugated bispinor, $\psi_{R}=\frac{1}{2}(1+\gamma^{5})\psi$ and $\psi_{L}=\frac{1}{2}(1-\gamma^{5})\psi$ are the right-handed and the left-handed fields accordingly, so that $\psi=\psi_{L}+\psi_{R}$. $A_{\mu}=(\varphi,-\mathbf{A}),A^{\mu}=(\varphi,\mathbf{A})$ are covariant and contravariant potential of the Abelian field, $F_{\mu\nu}=\partial_{\mu}A_{\nu}-\partial_{\nu}A_{\mu}$ is Faraday tensor.

If the scalar field $\varphi(x)$ is gauge transformed, then the potential $A_{\mu}$ and Dirac fields $\psi_{L,R}$ should be gauge transformed too:
\begin{equation}\label{1.3}
    \varphi(x)=\left|\varphi(x)\right|\rightarrow\left|\varphi(x)\right|e^{\mathrm{i}\theta(x)}
    \Rightarrow A_{\mu}\rightarrow A_{\mu}-\frac{1}{e}\partial_{\mu}\theta\Rightarrow\psi_{L,R}\rightarrow\psi_{L,R}e^{\mathrm{i}\theta(x)},
\end{equation}
so that the Lagrangian does not depend on the phase $\theta(x)$.

Scalar and Dirac field interact by Yukawa coupling, which is gauge invariant:
\begin{equation}\label{1.4}
    U_{\chi}=\chi\left|\varphi(x)\right|(\overline{\psi}_{L}\psi_{R}+\overline{\psi}_{R}\psi_{L})=
    m(\overline{\psi}_{L}\psi_{R}+\overline{\psi}_{R}\psi_{L})
    +\chi\phi(x)(\overline{\psi}_{L}\psi_{R}+\overline{\psi}_{R}\psi_{L}),
\end{equation}
where $m=\chi\varphi_{0}$ is mass of fermion, and $\phi(x)=\left|\varphi(x)\right|-\varphi_{0}$ is small deviation from equilibrium value $\varphi_{0}$. The potential (\ref{1.4}) is invariant under gauge transformations (\ref{1.3}). However, it should be noted, that the coupling (\ref{1.4}) has some artificial form. There are fundamental fields: scalar $\varphi$, spinor $\psi$ and gauge field $A_{\mu}$ (due to local gauge invariance), hence all term in Lagrangian should be combinations of these fields and operations on them, and their modules and phases appear as a result of operations in the original expressions, however Eq.(\ref{1.4}) contains the module $|\varphi|$ initially.

At the same time we can propose another expression of Yukawa coupling \cite{grig}:
\begin{equation}\label{1.5}
    U_{\chi}=\chi(\overline{\psi}_{L}\varphi\psi_{R}+\overline{\psi}_{R}\varphi^{+}\psi_{L})
    =\chi|\varphi|(\overline{\psi}_{L}\psi_{R}+\overline{\psi}_{R}\psi_{L})\cos\theta
    +\mathrm{i}\chi|\varphi|(\overline{\psi}_{L}\psi_{R}-\overline{\psi}_{R}\psi_{L})\sin\theta,
\end{equation}
At first glance it may seem that this expression is not invariant under gauge transformations due to presence of phase $\theta(x)$. In Eq.(\ref{1.5}) we can see that $\overline{\psi}_{L}\psi_{R}+\overline{\psi}_{R}\psi_{L}$ is a \emph{scalar}, but $\overline{\psi}_{L}\psi_{R}-\overline{\psi}_{R}\psi_{L}$ is a \emph{pseudoscalar}. At the same time, the phase $\theta$ is scalar, because the vector-potential $\mathbf{A}$ is vector: $\mathbf{r}\rightarrow -\mathbf{r}\Rightarrow\mathbf{A}\rightarrow -\mathbf{A}$, and gauge transformation is $\mathbf{A}\rightarrow \mathbf{A}+\frac{1}{e}\frac{\partial\theta}{\partial \mathbf{r}}$, then $\mathbf{r}\rightarrow -\mathbf{r}\Rightarrow\theta\rightarrow \theta$. Hence, in order to obtain the Dirac mass of a fermion we should choose the vacuum so, that $\theta_{0}=0$, that is $m_{D}=\chi\varphi_{0}\cos\theta_{0}=\chi\varphi_{0}$, where we consider global gauge transformation i.e with constant phase $\theta(x)=\theta_{0}=\mathrm{const}$. This choosing is not a selection of a specific phase for the following reasons.

\begin{figure}[ht]
\begin{center}
\includegraphics[width=15cm]{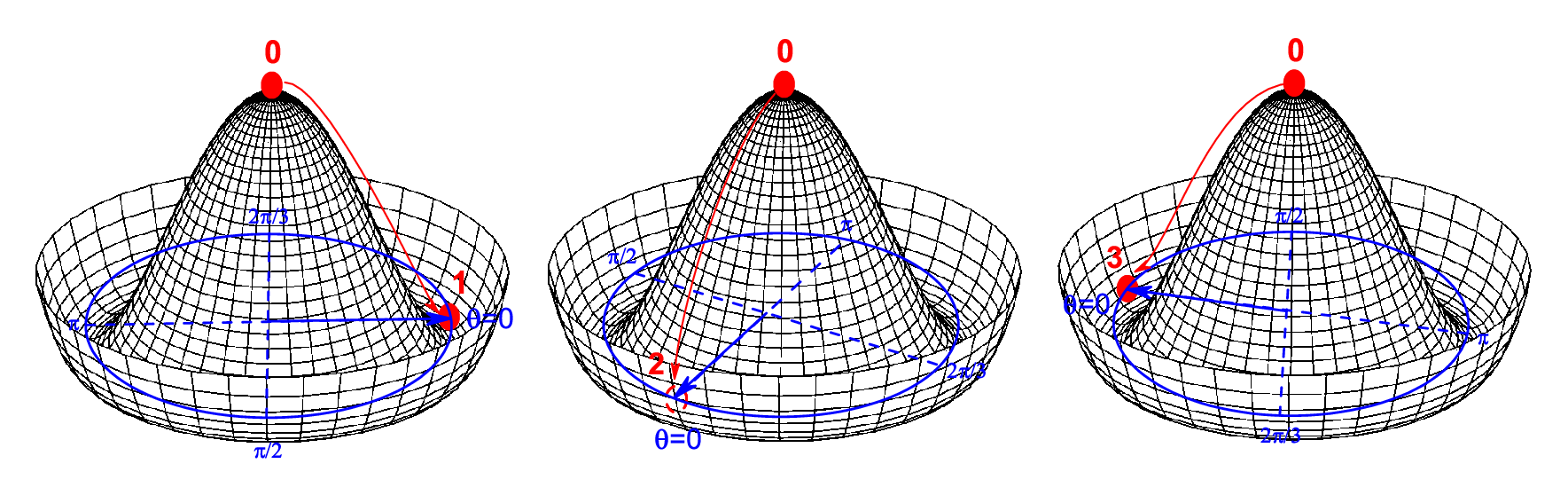}
\end{center}
\caption{the scalar field $\varphi$ from position 0 at the top of the hill rolls down into the valley, for example, to positions 1 or 2 or 3 having the equal energies. To each position we can assign an angular coordinate $\theta=0$.}
\label{Fig1}
\end{figure}

\begin{figure}[ht]
\begin{center}
\includegraphics[width=10cm]{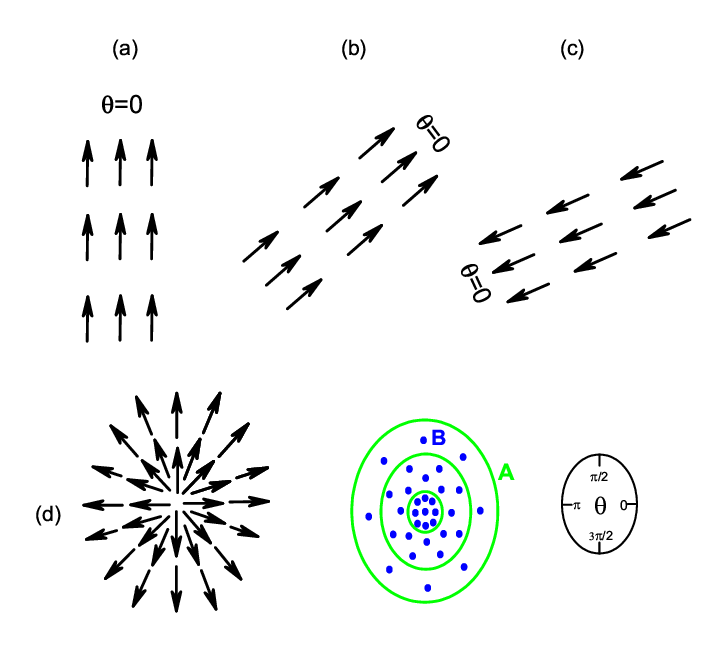}
\end{center}
\caption{configuration of scalar field $\varphi$ (arrows indicate the phase): (a,b,c) - in each point of space the scalar field has the same equilibrium phases $\varphi=|\varphi(x)|e^{\theta_{0}}$, and energies of the (a,b,c) are equal, then the phase value can be assigned to $\theta_{0}=0$, no matter in which phase $\theta$ the field has rolled. (d) - vortex, where there is a line with value of the scalar field $\varphi=0$ (in center of "hedgehog"), and magnetic field $\mathbf{B}$ (blue color) is directed along this axis, the concentric lines of vector-potential $\mathbf{A}$ are shown by green color.}
\label{Fig2}
\end{figure}

Let us consider the potential in a form of "mexican hat" in Fig.\ref{Fig1}. The field from the hill (position 0) rolls to any place in bottom of the valley (for example, positions 1, 2, 3). Due $U(1)$-symmetry of the potential all points in bottom of the valley are equivalent $V(\theta, |\varphi|=\varphi_{0})=\mathrm{const}$. Hence the phase $\theta$ can be always set as $\theta=0$ at any point, where the field has rolled down, that is, to set the begin for counting of angles from this point (we can suppose $\theta=\pi$ else, but then the Yukawa coupling constant must be $\chi<0$). Such procedure is caused by gravity. Let suppose $\theta\neq 0$, then mass of a fermion is sum of a scalar and a pseudoscalar $m=m_{s}+m_{a}$, i.e $\mathbf{r}\rightarrow -\mathbf{r}\Rightarrow m_{s}\rightarrow m_{s}, m_{a}\rightarrow -m_{a}$. Then from equation for gravitational potential $\Delta\varphi=4\pi G\rho$ we can see that the potential has scalar and pseudoscalar parts also $\varphi=\varphi_{s}+\varphi_{a}$. For simplicity, let us consider only the case of a pseudoscalar $\rho=\rho_{a}\Rightarrow\varphi=\varphi_{a}$ (that corresponds to $\theta=\pi/2$). Then the metric tensor $g_{\mu\nu}$ changes under the parity transformation as
\begin{equation}\label{1.6}
  ds^{2}=g_{\mu\nu}dx^{\mu}dx^{\nu}=\left(1+2\varphi\right)dt^{2}-\left(1-2\varphi\right)(dx^{2}+dy^{2}+dz^{2})
  \rightarrow\left(1-2\varphi\right)dt^{2}-\left(1+2\varphi\right)(dx^{2}+dy^{2}+dz^{2}).
\end{equation}
Thus, in general case we will have two metric tensors: a left $g_{\mu\nu}^{L}$ and a right $g_{\mu\nu}^{R}$, which are not equal $g_{\mu\nu}^{L}\neq g_{\mu\nu}^{R}$. Hence, we will have two Christoffel symbols: a left $\Gamma^{L\mu}_{\nu\kappa}$ and a right $\Gamma^{R\mu}_{\nu\kappa}$, which are not equal $\Gamma^{L\mu}_{\nu\kappa}\neq\Gamma^{R\mu}_{\nu\kappa}$. This means that we cannot make $\Gamma^{L\mu}_{\nu\kappa}=0$ and $\Gamma^{R\mu}_{\nu\kappa}=0$ simultaneously: choosing a coordinate system either as $x'^{\mu}=x^{\mu}+\frac{1}{2}\Gamma^{L\mu}_{\nu\kappa}x^{\nu}x^{\kappa}\Rightarrow \Gamma^{'L\mu}_{\nu\kappa}=0$ but then $\Gamma^{'R\mu}_{\nu\kappa}\neq 0$, or as $x'^{\mu}=x^{\mu}+\frac{1}{2}\Gamma^{R\mu}_{\nu\kappa}x^{\nu}x^{\kappa}\Rightarrow \Gamma^{'R\mu}_{\nu\kappa}=0$ but then $\Gamma^{'L\mu}_{\nu\kappa}\neq 0$. Thus, local inertial frame of reference does not exist, therefore the equivalence principle is violated in such a case. In addition, $\overline{\psi}_{L}\psi_{R}-\overline{\psi}_{R}\psi_{L}$ is not CP invariant. From the above it follows, no matter what point in the valley of potential in Fig.\ref{Fig1} the field rolls into, the equilibrium phase at that point is always assigned as $\theta_{0}=0$, that illustrated in Fig.\ref{Fig2}a,b,c.

\begin{figure}[ht]
\begin{center}
\includegraphics[width=8cm]{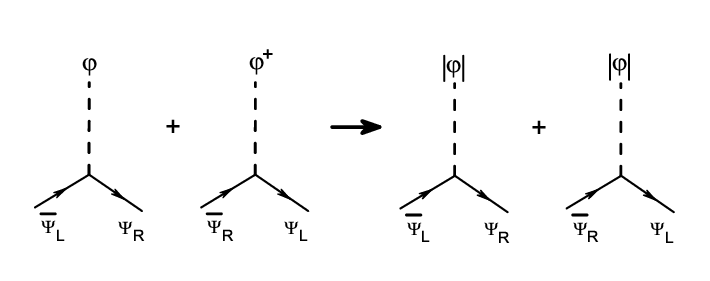}
\end{center}
\caption{transformation of vertexes of Yukawa coupling due to the absorption of phase oscillations $\theta(x)$ by the gauge field $A_{\mu}$ (Higgs effect): Dirac fields $\psi_{L,R}$ interact with module of the scalar field $|\varphi|$.}
\label{Fig3}
\end{figure}

\begin{figure}[ht]
\begin{center}
\includegraphics[width=8cm]{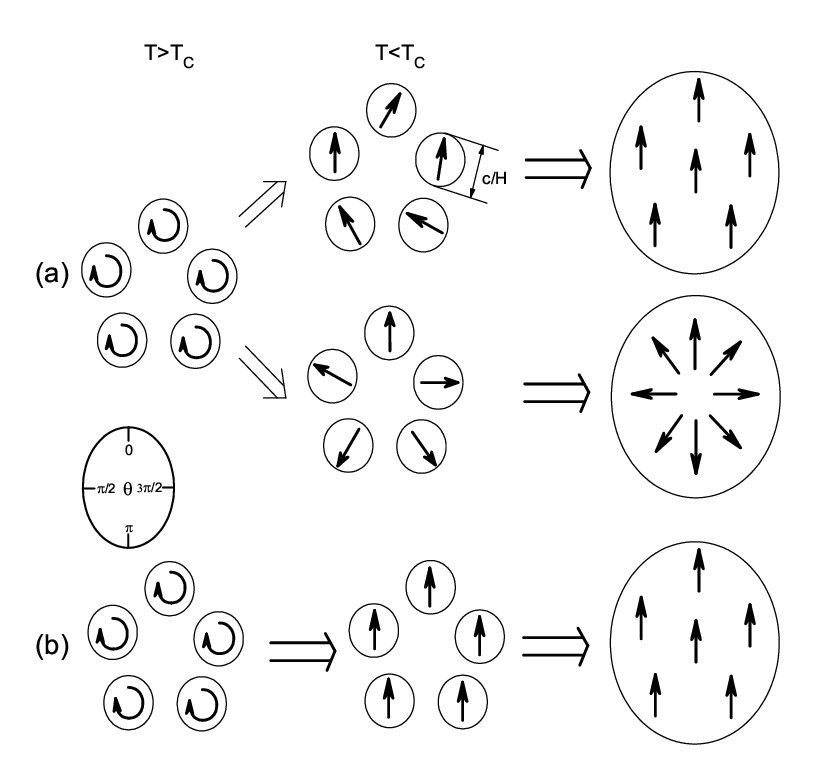}
\end{center}
\caption{distribution of phase of the scalar field $\varphi=|\varphi|e^{i\theta}$ throughout the entire space conditionally consisting of cells of the order cosmological horizon $\sim \frac{c}{H}$: (a) - evolution to either uniform distribution or to a topological defect as the horizon increases (the Hubble parameter $H$ decreases), (b) - application of the equivalence principle for Yukawa coupling results in the phase being assigned as $\theta=0$ in all cells, i.e Kibble–Zurek mechanism breaks down.}
\label{Fig4}
\end{figure}

Let us consider the local gauge transformation $|\varphi|e^{\theta(x)}$: despite the fact, that equilibrium phase is $\theta_{0}=0$, phase oscillations (Goldstone bosons) can takes place. The full interaction term has the form:
\begin{equation}\label{1.7}
    \mathcal{U}_{D}=\chi|\varphi|(\overline{\psi}_{L}\psi_{R}+\overline{\psi}_{R}\psi_{L})
    +\mathrm{i}\chi|\varphi|(\overline{\psi}_{L}\psi_{R}-\overline{\psi}_{R}\psi_{L})\theta,
\end{equation}
where $\theta=\theta(t,\mathbf{r})$ is small phase oscillations $|\theta|\ll 1$, that violates the gauge invariance of this coupling. Let us consider the currents corresponding to oscillations of the scalar field $J^{\mu}_{\varphi}=\mathrm{i}(\varphi^{+}\partial^{\mu}\varphi-\varphi\partial^{\mu}\varphi^{+})=-2|\varphi|^{2}\partial^{\mu}\theta$ and of the spinor field $J^{\mu}_{\psi}=\overline{\psi}\gamma^{\mu}\psi=\overline{\psi}_{L}\gamma^{\mu}\psi_{L}+\overline{\psi}_{R}\gamma^{\mu}\psi_{R}$. So, it is not difficulty to obtain:
\begin{eqnarray}
  \partial_{\mu}J^{\mu}_{\varphi}&=&\mathrm{i}\chi(\overline{\psi}_{L}\varphi\psi_{R}-\overline{\psi}_{R}\varphi^{+}\psi_{L}),\label{1.8a}\\
  \partial_{\mu}J^{\mu}_{\psi}&=&0.
\end{eqnarray}
That is, the current of fermions $J^{\mu}_{\psi}$ is conserved, but the current caused by the scalar field $J^{\mu}_{\varphi}$ is not conserved: in the vertexes $\overline{\psi}_{L}\varphi\psi_{R}$ and $\overline{\psi}_{R}\varphi^{+}\psi_{L}$, illustrated in Fig.\ref{Fig4}, the charge of the scalar field $\varphi$ disappears. Let us turn on the gauge field $A_{\mu}$, then the current $J^{\mu}_{\varphi}$ takes the form:
\begin{equation}\label{1.9}
  J^{\mu}_{\varphi}=\mathrm{i}\left[\varphi^{+}(\partial^{\mu}+\mathrm{i}eA^{\mu})\varphi-\varphi(\partial^{\mu}-\mathrm{i}eA^{\mu})\varphi^{+}\right]=
  -2|\varphi|^{2}\left(\partial^{\mu}\theta+eA^{\mu})\right)=-2|\varphi|^{2}eA^{\mu},
\end{equation}
where we have used the gauge transformation $A^{\mu}\rightarrow A^{\mu}-\frac{1}{e}\partial^{\mu}\theta$. Then,
\begin{equation}\label{1.10}
  \partial_{\mu}J^{\mu}_{\varphi}=-2eA^{\mu}\partial_{\mu}|\varphi|^{2}-2e|\varphi|^{2}\partial_{\mu}A^{\mu}=
  -2e\left(-\frac{2e^{2}}{b}A_{\mu}A^{\mu}+|\varphi|^{2}\right)\partial_{\mu}A^{\mu}=0,
\end{equation}
where we used the 4-dimensional transverseness $\partial_{\mu}A^{\mu}=0$ (the Lorentz gauge becomes mandatory for massive gauge field), and potential of the scalar field at presence of gauge field of $a|\varphi|^{2}+\frac{b}{2}|\varphi|^{4}+e^{2}A_{\mu}A^{\mu}|\varphi|^{2}\Rightarrow|\varphi|^{2}=-\frac{a}{b}-\frac{e^{2}}{b}A_{\mu}A^{\mu}
\Rightarrow\partial_{\mu}|\varphi|^{2}=-\frac{2e^{2}}{b}A_{\mu}(\partial_{\mu}A^{\mu})$. Thus, the current of Dirac field $J^{\mu}_{\psi}$ and the current of scalar field $J^{\mu}_{\varphi}$ are conserved separately: $\partial_{\mu}J^{\mu}_{\varphi}=0$ and $\partial_{\mu}J^{\mu}_{\psi}=0$. This means, that in the vertexes $\overline{\psi}_{L}\varphi\psi_{R}$ and $\overline{\psi}_{R}\varphi^{+}\psi_{L}$ the charge flows along the fermion lines, while the scalar field lines $\varphi,\varphi^{+}$ carry no charge due to Higgs mechanism: the phase oscillations $\theta(t,\mathbf{r})$ are absorbed by the gauge fields $A_{\mu}$, and the \emph{real} field $|\varphi|$ does not carry any charge. That is the transformation takes place: $\varphi,\varphi^{+}\rightarrow|\varphi|,|\varphi|$ in the vertexes of Yukawa coupling, that illustrated in Fig.\ref{Fig3}. Hence, the Lagrangian is transformed at spontaneous $U(1)$ symmetry breaking as
\begin{equation}\label{1.11}
  (\partial_{\mu}+\mathrm{i}eA_{\mu})\varphi(\partial^{\mu}-\mathrm{i}eA^{\mu})\varphi^{+}
  -\chi(\overline{\psi}_{L}\varphi\psi_{R}+\overline{\psi}_{R}\varphi^{+}\psi_{L})
  \rightarrow\partial_{\mu}|\varphi|\partial^{\mu}|\varphi|
  +e^{2}|\varphi|^{2}A_{\mu}A^{\mu}-\chi|\varphi|(\overline{\psi}_{L}\psi_{R}+\overline{\psi}_{R}\psi_{L}),\quad \theta=0,
\end{equation}
where the choice of vacuum $\theta=0$ is caused by the principle of equivalence.


Thus, in each point of space the scalar field $\varphi=|\varphi(x)|e^{\theta_{0}}$ has the same equilibrium phases, which must be set as $\theta_{0}=0$, no matter in which direction (by angle $\theta$) the field has rolled down - Fig.\ref{Fig2}(a,b,c). At the same time, in topological defects as, for example, vortex (string) illustrated in Fig.\ref{Fig2}(d), there is a space region where the scalar field did not roll into the valley, i.e $\varphi=0$, which is a line or a ring, and around the defect the phase changes within the limits $\theta=[0,2\pi]$. Along the string the fermionic string superconductivity take place: fermions are massless $(\partial_{t}+\partial_{z})\psi(z,t)=0$ along the string ($Oz$ axis) and the fermionic current in electric field $E$ is $J=\frac{q^{2}Et}{2\pi}$ \cite{vilen,rubakov}.


For topological defects to exist, a mechanism for their formation is necessary in cosmological processes. So, according to Kibble-Zurek mechanism \cite{kibble1,kibble2,vilen1,vilen2,zurek1,zurek2}, at high temperatures $T>T_{c}$ the phase has no fixed value (it is fluctuating rapidly), but after spontaneously breakdown of a global symmetry (at $T<T_{c}$) the phase become homogeneous $\theta=\mathrm{const}$ within the correlation length of order of the cosmological horizon $l_{\mathrm{cor}}\sim c/H(t)$, where $H$ if the Hubble parameter as function of time, at the same time in other cells the values of phase can be other because regions of space, separated by a distance greater than the cosmological horizon, do not have time to correlate with each other - Fig.\ref{Fig4}a. With the Hubble parameter is decreasing, the cosmological horizon increases, then the initial random phases are correlated either into a uniform distribution or into a topological defect. However application of the equivalence principle for Yukawa coupling results in the phase being assigned as $\theta=0$ in all cells. Then, the uniform phase distribution is established throughout space, regardless of the magnitude of the cosmological horizon - Fig.\ref{Fig4}b. As a result, the Kibble-Zurek mechanism breaks down and topological defects cannot form.


At the same time, there is mechanism for formation of topological defects in system with a local gauge symmetry \cite{hind,rajan}. Although the magnetic flux is zero on the average $\langle\Phi\rangle=0$, its thermal fluctuations exist, i.e there are regions some characteristic size $\xi$, where $\Phi\neq 0$. When the system enters the broken phase, this flux is freezed, and then the field condensate rearranges the field configuration to minimize the magnetic flux and the energy associated with it, so that the magnetic flux is pressed into the center of a topological defect - a vortex. 

\section{Gauge $SU(2)\times U(1)$ symmetry}\label{SU2}

Unlike the toy model based on $U(1)$ gauge symmetry, the model based on the symmetry $SU(2)_{I}\times U(1)_{Y}$ spontaneous breaking to $U(1)_{Q}$ corresponds to real physics: the electro-weak interaction. Let $\Psi$ is isospinor:
\begin{equation}\label{2.1}
    \Psi(x)=e^{\mathrm{i}\theta(x)}e^{\mathrm{i}\frac{\vec{\tau}}{2}\vec{\vartheta}(x)}\left(\begin{array}{c}
           0 \\
           \varphi(x) \\
         \end{array}\right),
\end{equation}
where $\varphi>0$ is real,
\begin{equation}\label{2.2}
  e^{\mathrm{i}\frac{\vec{\tau}}{2}\vec{\vartheta}}\equiv\left(\cos\frac{\vartheta}{2}+\mathrm{i}(\vec{n}\vec{\tau})\sin\frac{\vartheta}{2}\right),
\end{equation}
where $\vec{\tau}=(\tau_{x},\tau_{y},\tau_{z})$ is a vector consisting of Pauli matrices, $\vec{\vartheta}=\vec{n}\vartheta$, where $\vec{n}$ is an unit vector in the direction of the axis around which the rotation is made in the isospace. Thus, we assign the third projection of isospin as $I_{z}=-\frac{1}{2}$ to the scalar field $\varphi$, then hypercharge $Y=1$ so, that the electrical charge is $Q=I_{z}+\frac{Y}{2}=0$.

Fermions are described with spinor field which are isodublet and isosinglet accordingly:
\begin{equation}\label{2.3}
  \psi_{L}=\left(\begin{array}{c}
    \nu_{L} \\
    e_{L} \\
    \end{array}\right),\quad\psi_{R}=e_{R},
\end{equation}
where $e_{L}$ is a left-handed electron with $I_{z}=-\frac{1}{2}, Y=-1$, $e_{R}$ is a right-handed electron with $I_{z}=0, Y=-2$, $\nu_{L}$ is left-handed neutrino with $I_{z}=\frac{1}{2}, Y=-1$.

Corresponding Lagrangian $\mathcal{L}$ is:
\begin{eqnarray}\label{2.4}
    \mathcal{L}=L_{\Psi}+L_{\psi}+L_{\chi}+L_{A,B}&\equiv&D_{\mu}\Psi\left(D^{\mu}\Psi\right)^{+}-a\Psi\Psi^{+}-\frac{b}{2}\left(\Psi\Psi^{+}\right)^{2}\nonumber\\
    &+&\mathrm{i}\overline{\psi}_{L}\gamma^{\mu}D_{\mu}^{\vec{A},B}\psi_{L}+\mathrm{i}\overline{\psi}_{R}\gamma^{\mu}D_{\mu}^{B}\psi_{R}\nonumber\\
    &-&\chi\left[\overline{\psi}_{L}\Psi\psi_{R}+\overline{\psi}_{R}\Psi^{+}\psi_{L}\right]\nonumber\\
    &-&\frac{1}{16\pi}G_{\mu\nu}G^{\mu\nu}-\frac{1}{16\pi}\vec{F}_{\mu\nu}\vec{F}^{\mu\nu},
\end{eqnarray}
where
\begin{eqnarray}
    D_{\mu}&\equiv&\partial_{\mu}-\mathrm{i}g\frac{\vec{\tau}}{2}\vec{A}_{\mu}-\mathrm{i}\frac{f}{2}B_{\mu},\label{2.5a}\\
    D^{\vec{A},B}_{\mu}&\equiv&\partial_{\mu}
    -\mathrm{i}g\frac{\vec{\tau}}{2}\vec{A}_{\mu}+\mathrm{i}\frac{f}{2}B_{\mu},\label{2.5b}\\
    D^{B}_{\mu}&\equiv&\partial_{\mu}+\mathrm{i}fB_{\mu}\label{2.5c}
\end{eqnarray}
are covariant derivations: Abelian field $B_{\mu}$ corresponds to the local gauge $U(1)$ symmetry, and the nonabelian field $\vec{A}_{\mu}$ corresponds to the local gauge $SU(2)$ symmetry, $f$ and $g$ are corresponding coupling constants,
\begin{equation}\label{2.6}
  \vec{F}_{\mu\nu}=\partial_{\mu}\vec{A}_{\nu}-\partial_{\nu}\vec{A}_{\mu}+g\left[\vec{A}_{\mu}\times\vec{A}_{\nu}\right],\quad
  G_{\mu\nu}=\partial_{\mu}B_{\nu}-\partial_{\nu}B_{\mu}
\end{equation}
are tensors of the Yang–Mills field $\vec{A}_{\mu}$ and Abelian field $B_{\mu}$ accordingly. The gauge field are transformed at gauge transformation of isospinor field (\ref{2.1})
\begin{equation}\label{2.7}
  \Psi\rightarrow e^{\mathrm{i}\theta(x)}e^{\mathrm{i}\frac{\vec{\tau}}{2}\vec{\vartheta}(x)}\Psi
\end{equation}
as
\begin{eqnarray}
    \vec{A}_{\mu}&\rightarrow&\vec{A}_{\mu}+\frac{1}{g}\partial_{\mu}\vec{\vartheta}-\left[\vec{\vartheta}\times\vec{A}_{\mu}\right],\label{2.8a}\\
    B_{\mu}&\rightarrow&B_{\mu}+\frac{2}{f}\partial_{\mu}\theta\label{2.8b}.
\end{eqnarray}
The Dirac fields are transformed at gauge transformation (\ref{2.7}) as:
\begin{equation}\label{2.9}
  \psi_{L}\rightarrow e^{-\mathrm{i}\theta(x)}e^{\mathrm{i}\frac{\vec{\tau}}{2}\vec{\vartheta}(x)}\psi_{L},\quad
  \psi_{R}\rightarrow e^{-2\mathrm{i}\theta(x)}\psi_{R}.
\end{equation}
Then Yukawa coupling is invariant under simultaneous gauge transformations (\ref{2.7}) and (\ref{2.9}):
\begin{equation}\label{2.10}
  \left[\overline{\psi}_{L}\Psi\psi_{R}+\overline{\psi}_{R}\Psi^{+}\psi_{L}\right]
  \rightarrow\left[\overline{\psi}_{L}\Psi\psi_{R}+\overline{\psi}_{R}\Psi^{+}\psi_{L}\right].
\end{equation}

Nevertheless, synchronous rotation of isospinor $\Psi$ and Dirac fields $\psi_{L},\psi_{R}$ must be provided by some physical mechanism. Otherwise, if we rotate these fields "by hand", it will not be symmetry, but a constraint imposed on the system. Such a mechanism is the interaction with gauge fields $\vec{A}_{\mu}$ and $B_{\mu}$: if we rotate the isospinor field $\Psi$ by gauge transformation (\ref{2.7}), then, according to the transformations (\ref{2.8a},\ref{2.8b}), we generate the gauge fields as
\begin{eqnarray}
    \vec{A}_{\mu}&=&\frac{1}{g}\partial_{\mu}\vec{\vartheta},\label{2.11a}\\
    B_{\mu}&=&\frac{2}{f}\partial_{\mu}\theta\label{2.11b}.
\end{eqnarray}
In turn, from the Lagrangian (\ref{2.4}) we can see that the Dirac fields also interact with the gauge fields, that is described by the term $L_{\psi}$. This means that the Dirac fields should also rotate in response to appearance of the gauge fields:
\begin{eqnarray}
    \overline{\psi}_{L}\gamma^{\mu}\partial_{\mu}\psi_{L}&=&
    e^{\mathrm{i}\theta(x)}e^{-\mathrm{i}\frac{\vec{\tau}}{2}\vec{\vartheta}(x)}\overline{\psi}_{L}\gamma^{\mu}\left(\partial_{\mu}
    -\mathrm{i}\frac{\vec{\tau}}{2}\partial_{\mu}\vec{\vartheta}+\mathrm{i}\partial_{\mu}\theta\right)
    e^{-\mathrm{i}\theta(x)}e^{\mathrm{i}\frac{\vec{\tau}}{2}\vec{\vartheta}(x)}\psi_{L},\label{2.12a}\\
    \overline{\psi}_{R}\gamma^{\mu}\partial_{\mu}\psi_{R}&=&
    e^{2\mathrm{i}\theta(x)}\overline{\psi}_{R}\gamma^{\mu}\left(\partial_{\mu}
    +2\mathrm{i}\partial_{\mu}\theta\right)e^{-2\mathrm{i}\theta(x)}\psi_{R}\label{2.12b}.
\end{eqnarray}

\begin{figure}[ht]
\begin{center}
\includegraphics[width=10cm]{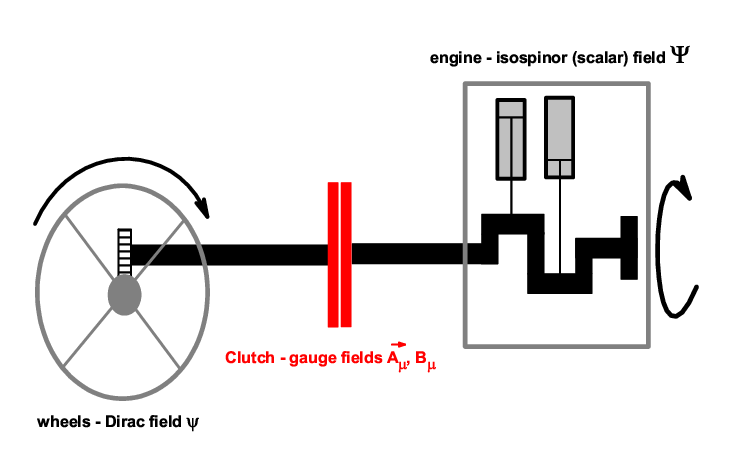}
\end{center}
\caption{the mechanical analogy of the transmission of rotation from isospinor (scalar) field $\Psi$ to Dirac field $\psi$ via the gauge fields $\vec{A}_{\mu},B_{\mu}$ and vice versa. Rotation from the engine is transmitted to the wheels via the clutch - gauge fields. And vice versa: you can rotate the wheels and this rotation will be transmitted to the engine shaft via the clutch. However, in the case of rotation by the global gauge transformations, the clutch is not cuddled and, as a consequence, the backlash in rotation occurs.}
\label{Fig5}
\end{figure}

However, let us consider the \emph{global} gauge transformation, that is where $\theta=\mathrm{const}$ and $\overrightarrow{\vartheta}=\mathrm{const}$. Then, from Eqs.(\ref{2.11a},\ref{2.11b}) we can see, that the gauge field are not generated: $B_{\mu}=0,\vec{A}_{\mu}=0$. This means that the isospinor field $\Psi$, rotating, cannot clatch the Dirac field $\psi$ and vice versa:
\begin{equation}\label{2.13}
  \delta\Psi\nLeftrightarrow(B_{\mu},\vec{A}_{\mu})\nLeftrightarrow(\delta\psi_{L},\delta\psi_{R}),\quad\texttt{ if } \theta=\mathrm{const},\overrightarrow{\vartheta}=\mathrm{const}.
\end{equation}
This expresses the fact of short-range action in nature: interaction of charges with fields occures, but not interaction of charges with each other (i.e the long-range action). In Fig.\ref{Fig5} the mechanical analogy of above-described transmission of rotation from isospinor (scalar) field $\Psi$ to Dirac field $\psi$ via the gauge fields $\vec{A}_{\mu},B_{\mu}$ and vice versa is proposed.

Let us, for example, turn the field $\Psi$ as $\Psi\rightarrow e^{\mathrm{i}\theta}e^{\mathrm{i}\frac{\vec{\tau}}{2}\vartheta}\Psi$ but we do not turn the fields $e_{L},e_{R}$, then the mass term for an electron will be:
\begin{eqnarray}\label{2.14}
  &&\chi\varphi_{0}\left[\overline{e}_{L}e^{\mathrm{i}\theta}e_{R}+\overline{e}_{R}e^{-\mathrm{i}\theta}e_{L}\right]\cos\frac{\vartheta}{2}
  -\mathrm{i}\chi\varphi_{0}n_{z}\left[\overline{e}_{L}e^{\mathrm{i}\theta}e_{R}-\overline{e}_{R}e^{-\mathrm{i}\theta}e_{L}\right]
  \sin\frac{\vartheta}{2}\nonumber\\
  &&=\chi\varphi_{0}\left[\overline{e}_{L}e_{R}+\overline{e}_{R}e_{L}\right]\cos\theta\cos\frac{\vartheta}{2}
  +\chi\varphi_{0}n_{z}\left[\overline{e}_{L}e_{R}+\overline{e}_{R}e_{L}\right]\sin\theta\sin\frac{\vartheta}{2}\nonumber\\
  &&-\mathrm{i}\chi\varphi_{0}\left[\overline{e}_{L}e_{R}-\overline{e}_{R}e_{L}\right]\sin\theta\cos\frac{\vartheta}{2}
  -\mathrm{i}\chi\varphi_{0}n_{z}\left[\overline{e}_{L}e_{R}-\overline{e}_{R}e_{L}\right]
  \cos\theta\sin\frac{\vartheta}{2},
\end{eqnarray}
where supposed $n_{x}=n_{y}=0,n_{z}=\pm 1$. The mass term (\ref{2.14}) is in agreement with the equivalence principle in gravity if ($\chi>0$, $n_{z}=1$):
\begin{equation}\label{2.15}
  \begin{array}{cc}
    \mathrm{i}&\theta=0, \vartheta=0 \\
    \mathrm{ii}&\theta=\frac{\pi}{2}, \vartheta=\pi \\
    \mathrm{iii}&\theta=\pi, \vartheta=2\pi \\
    \mathrm{iiii}&\theta=\frac{3\pi}{2}, \vartheta=3\pi \\
  \end{array}
\end{equation}
Either we can turn the Dirac fields $\psi_{L}\rightarrow e^{-\mathrm{i}\theta}e^{\mathrm{i}\frac{\vec{\tau}}{2}\vartheta}\psi_{L}$ and $e_{R}\rightarrow e^{-2\mathrm{i}\theta}e_{R}$ but we do not turn the field $\Psi$, then we will obtain Eq.(\ref{2.15}) for phases of Dirac fields. Thus, we have four vacuums (\ref{2.15}). Regions with different vacuums should be separated by a domain wall. Within the wall we will have very strong gauge fields (\ref{2.11a},\ref{2.11b}) $A\sim\frac{1}{g}\frac{\Delta\vartheta}{\xi}\sim\frac{m_{H}}{g}$, $B\sim\frac{2}{f}\frac{\Delta\theta}{\xi}\sim\frac{m_{H}}{f}$, where $\xi\propto 1/m_{H}$ is a coherence length, $m_{H}$ is mass of Higgs boson. Such a wall contains energy with surface density $m_{H}^{3}\left(\frac{1}{f^{2}}+\frac{1}{g^{2}}\right)$. The formation of a domain wall between the vacuums should occur via the Kibble-Zurek mechanism. Thus, unlike simple groups $SU(n)$, in the case of combined group, as $SU(2)\times U(1)$, the different vacuums can occur, which are separated be domain walls.

In single-band systems the assignation (\ref{2.15}) can be always chosen. However, as demonstrated in Ref.\cite{grig}, in multi-band systems we should choose the equilibrium phases $\vec{\vartheta}^{0}_{1,2,3},\theta^{0}_{1,2,3}$ are equal for all condensates, for examples, as $\vec{\vartheta}^{0}_{1}=\vec{\vartheta}^{0}_{2}=\vec{\vartheta}^{0}_{3}=0,\theta^{0}_{1}=\theta^{0}_{2}=\theta^{0}_{3}=0$, that possible only when parameter of interband interaction is $\epsilon<0$.

It should be noted, that the Yukawa coupling by itself is not obliged to coordinate the global phases $\theta,\vartheta$ of the fields $\Psi$ and $\psi$ so that $\overline{\psi}_{L}\Psi\psi_{R}+\overline{\psi}_{R}\Psi^{+}\psi_{L}\rightarrow \overline{\psi}_{L}\Psi\psi_{R}+\overline{\psi}_{R}\Psi^{+}\psi_{L}$. Let us, for example, turn the field $\Psi$ as $\Psi\rightarrow e^{\mathrm{i}\theta}\Psi$ but we do not turn the fields $\psi_{L},e_{R}$ (or vice versa), then we obtain:
\begin{eqnarray}\label{2.16}
  \overline{\psi}_{L}\Psi\psi_{R}+\overline{\psi}_{R}\Psi^{+}\psi_{L}&\rightarrow&
  \varphi\left[\overline{e}_{L}e^{\mathrm{i}\theta}e_{R}+\overline{e}_{R}e^{-\mathrm{i}\theta}e_{L}\right]\nonumber\\
  &=&\varphi\left[\overline{e}_{L}e_{R}+\overline{e}_{R}e_{L}\right]\cos\theta
  +\mathrm{i}\varphi\left[\overline{e}_{L}e_{R}-\overline{e}_{R}e_{L}\right]\sin\theta
  =\varphi \overline{e}e\cos\theta+\mathrm{i}\varphi\overline{e}\gamma^{5}e\sin\theta.
\end{eqnarray}
Then making $P$, $CP$ and $CPT$ transformations we have:
\begin{equation}\label{2.17}
  \begin{array}{cc}
    P: & \overline{e}e\rightarrow \overline{e}e, \quad \mathrm{i}\overline{e}\gamma^{5}e\rightarrow -\mathrm{i}\overline{e}\gamma^{5}e \\
    CP: & \overline{e}e\rightarrow \overline{e}e, \quad \mathrm{i}\overline{e}\gamma^{5}e\rightarrow -\mathrm{i}\overline{e}\gamma^{5}e \\
    CPT: & \overline{e}e\rightarrow \overline{e}e, \quad \mathrm{i}\overline{e}\gamma^{5}e\rightarrow \mathrm{i}\overline{e}\gamma^{5}e
  \end{array}
\end{equation}
We can see that Yukawa coupling is $CPT$ invariant even in the above case of phase misalignment of $\Psi$ and $\psi$. This means that this coupling does not have to coordinate phases, that is to be invariant under the global gauge transformation. To coordinate phases, space-time symmetry is necessary - the equivalence principle by the above-described method.


\section{Strong CP problem}\label{CP}

Let us consider local gauge color $SU(3)_{c}$ symmetry. Then we have gauge field $A_{\mu}^{a}$, where $a=1\ldots 8$ and Dirac fields $\psi^{\alpha}$, where index $\alpha$ runs through six quarks $\alpha=u,d,c,s,t,b$. QCD Lagrangian is:
\begin{equation}\label{3.1}
    \mathcal{L}=\mathrm{i}\overline{\psi}_{L}^{\alpha}\gamma^{\mu}D_{\mu}\psi_{L}^{\alpha}
    +\mathrm{i}\overline{\psi}_{R}^{\alpha}\gamma^{\mu}D_{\mu}\psi_{R}^{\alpha}
    -m_{\alpha}\left[\overline{\psi}_{L}^{\alpha}e^{2\mathrm{i}\beta}\psi_{R}^{\alpha}
    +\overline{\psi}_{R}^{\alpha}e^{-2\mathrm{i}\beta}\psi_{L}^{\alpha}\right]
    -\frac{1}{4}G_{\mu\nu}^{a}G^{\mu\nu a}-\frac{\alpha_{s}}{8\pi}\theta_{0}G_{\mu\nu}^{a}\widetilde{G}^{\mu\nu a},
\end{equation}
where
\begin{eqnarray}
    G_{\mu\nu}^{a}&=&\partial_{\mu}A_{\nu}^{a}-\partial_{\nu}A_{\mu}^{a}+gf^{abc}A_{\mu}^{b}A_{\mu}^{c}\label{3.2}\nonumber\\
    \widetilde{G}^{\mu\nu a}&=&\frac{1}{2}\epsilon^{\mu\nu\lambda\rho}\widetilde{G}^{a}_{\mu\nu}\label{3.3}\nonumber\\
    D_{\mu}\psi&=&\partial_{\mu}\psi-\mathrm{i}g\frac{\lambda^{a}}{2}A_{\mu}^{a}\psi.\label{3.4}
\end{eqnarray}
Here scalar parameters $\beta$ and $\theta_{0}$ make the Lagrangian CP non-invariant, because the Yukawa coupling expression is
\begin{equation}\label{3.7}
  m\left[\overline{\psi}_{L}e^{2\mathrm{i}\beta}\psi_{R} +\overline{\psi}_{R}e^{-2\mathrm{i}\beta}\psi_{L}\right]=m\left[\overline{\psi}_{L}\psi_{R}+\overline{\psi}_{R}\psi_{L}\right]\cos\beta
+m\left[\overline{\psi}_{L}\psi_{R}-\overline{\psi}_{R}\psi_{L}\right]\sin\beta,
\end{equation}
where $\overline{\psi}_{L}\psi_{R}-\overline{\psi}_{R}\psi_{L}$ is a pseudoscalar, and
\begin{equation}\label{3.7a}
  G_{\mu\nu}^{a}\widetilde{G}^{\mu\nu a}\propto \mathbf{E}_{a}\mathbf{H}_{a}
\end{equation}
is a pseudoscalar also. At the same time, these terms are CPT-invariant that allows their existence. If we transform quarks fields by global axial transformations ($\beta=\mathrm{const}$):
\begin{equation}\label{3.5}
  \begin{array}{c}
    \psi_{L}^{\alpha}\rightarrow e^{\mathrm{i}\beta}\psi_{L}^{\alpha} \\
    \psi_{R}^{\alpha}\rightarrow e^{-\mathrm{i}\beta}\psi_{R}^{\alpha}
  \end{array},
\end{equation}
then $\overline{\psi}_{L}^{\alpha}e^{2\mathrm{i}\beta}\psi_{R}^{\alpha}+\overline{\psi}_{R}^{\alpha}e^{-2\mathrm{i}\beta}\psi_{L}^{\alpha}\rightarrow \overline{\psi}_{L}^{\alpha}\psi_{R}^{\alpha}+\overline{\psi}_{R}^{\alpha}\psi_{L}^{\alpha}$, but
\begin{equation}\label{3.6}
  \theta\rightarrow\theta+2N_{f}\beta,
\end{equation}
where $N_{f}$ is number of types of quarks. Thus we cannot get rid of the parameters $\beta$ and $\theta_{0}$ simultaneously, hence the strong interaction is CP non-invariant essentially. However experimental data limits the CP phase as $|\theta_{0}+2N_{f}\beta|<0.3\cdot 10^{-9}$, that which is much less than the CP-violation due to quark mixing by CKM matrix. To explain the absence of such a CP phase, the hypothetical particles - axions were proposed: there is an additional dynamic field $\theta(x)$ which interacts with quark field $m_{\alpha}\overline{\psi}_{L}^{\alpha}e^{2\mathrm{i}\beta+i\theta(x)}\psi_{R}^{\alpha}+h.c.$, and it provides $\theta_{0}+2N_{f}\beta+\langle\theta(x)\rangle N_{f}=0$. However the axions have not yet been detected, moreover, the existence of axions would lead to additional energy emission from stars and white dwarfs, which contradicts observational data. In resent research \cite{infirri} it is reported that no signal from axion has been observed, viable hadronic axion models are ruled out in a post-inflationary region $m>40\mu eV$. 

Above, we could see that the Yukawa coupling expression (\ref{3.7}) turns into
\begin{equation}\label{3.8}
 \beta=0\Rightarrow m\left[\overline{\psi}_{L}\psi_{R}+\overline{\psi}_{R}\psi_{L}\right]
\end{equation}
due to described above mechanism caused by the equivalence principle. Then we have following terms of Lagrangian:
\begin{equation}\label{3.9}
    -m_{\alpha}\left[\overline{\psi}_{L}^{\alpha}\psi_{R}^{\alpha}+\overline{\psi}_{R}^{\alpha}\psi_{L}^{\alpha}\right]
    -\frac{\alpha_{s}}{8\pi}\theta_{0} G_{\mu\nu}^{a}\widetilde{G}^{\mu\nu a}.
\end{equation}
Shifting the phase $\theta\rightarrow\theta-\theta=0$ we make turns of the Dirac fields as
\begin{equation}\label{3.9a}
    \psi_{L}^{\alpha}\rightarrow \exp\left(-\mathrm{i}\frac{\theta_{0}}{2N_{f}}\right)\psi_{L}^{\alpha},\quad
    \psi_{R}^{\alpha}\rightarrow \exp\left(\mathrm{i}\frac{\theta_{0}}{2N_{f}}\right)\psi_{R}^{\alpha},
\end{equation}
so that we obtain instead of Eq.(\ref{3.9}):
\begin{equation}\label{3.10}
m_{\alpha}\left[\overline{\psi}_{L}^{\alpha}\psi_{R}^{\alpha}+\overline{\psi}_{R}^{\alpha}\psi_{L}^{\alpha}\right]
    +\frac{\alpha_{s}}{8\pi}\theta_{0} G_{\mu\nu}^{a}\widetilde{G}^{\mu\nu a}\rightarrow
    m_{\alpha}\left[\overline{\psi}_{L}^{\alpha}\exp\left(\mathrm{i}\frac{\theta_{0}}{N_{f}}\right)\psi_{R}^{\alpha}
    +\overline{\psi}_{R}^{\alpha}\exp\left(-\mathrm{i}\frac{\theta_{0}}{N_{f}}\right)\psi_{L}^{\alpha}\right].
\end{equation}
However, the accounting of the equivalence principle leads to $\theta_{0}=0$, that is the gravity provides CP-invariance of QCD Lagrangian:
\begin{equation}\label{3.11}
    \mathcal{L}=\mathrm{i}\overline{\psi}_{L}^{\alpha}\gamma^{\mu}D_{\mu}\psi_{L}^{\alpha}
    +\mathrm{i}\overline{\psi}_{R}^{\alpha}\gamma^{\mu}D_{\mu}\psi_{R}^{\alpha}
    -m_{\alpha}\left[\overline{\psi}_{L}^{\alpha}\psi_{R}^{\alpha}+\overline{\psi}_{R}^{\alpha}\psi_{L}^{\alpha}\right]
    -\frac{1}{4}G_{\mu\nu}^{a}G^{\mu\nu a}.
\end{equation}
Thus, the strong CP problem is solved, hence the axion hypothesis is redundant.

\section{Results}\label{results}

In this work we investigate the effect of Yukawa coupling between scalar and fermion fields of type $\bar{\psi}_{L}\varphi\psi_{R}+\bar{\psi}_{R}\varphi^{+}\psi_{L}$ and the principle of equivalence in gravity on formation of topological defects and strong CP non-invariance. Corresponding results are:
\begin{itemize}
  \item The Yukawa coupling is non-invariant under global gauge transformations by groups $U(1), SU(2)$ and so on due to the backlash in clutch between phase rotations of scalar field and Dirac fields. At the same time, for local gauge transformations corresponding clutch is being carried out by gauge fields. As a result, Dirac masses depend on global phase and they are non P and CP invariant. However the equivalence principle in gravity assigns the phases $\theta=0$, $\vec{\vartheta}=0$, and so on, for scalar (isospinor) fields and Dirac fields in each point in space due to gauge invariance of the potential of scalar field $U(\varphi)$, so that P and CP invariance are restored. Small oscillations of phase are absorbed by gauge fields according to Higgs mechanism.
  \item As a result, the Kibble-Zurek mechanism breaks down and topological defects cannot form in system with a global gauge symmetry only. The uniform phase distribution is established throughout space, regardless of the magnitude of the cosmological horizon. At the same time, formation of topological defects in system with a local gauge symmetry is possible.
  \item At the same time, unlike simple groups $SU(n)$, in the case of combined group, as $SU(2)\times U(1)$, the different vacuums can occur (determined by different equilibrium phases $\theta,\vec{\vartheta}$), which are separated by domain walls. The formation of a domain wall between the vacuums should occur via the Kibble-Zurek mechanism.
  \item Due to the equivalence principle, the global chiral phase of fermions $\beta$ is assigned $\beta=0$. The angle $\theta_{0}$ can be assigned $\theta_{0}=0$ also, because it can be transformed to the chiral phase $\theta_{0}\rightarrow\exp\left(\pm\frac{\theta_{0}}{2N_{f}}\right)$ in the mass term. Thus, Lagrangian remains CP invariant. This mechanism makes the axion hypothesis needless.
\end{itemize}
Thus we can see, the gauge symmetry and the axial symmetry of scalar (isospinor) and Dirac fields are not formal mathematical symmetries, but they are dynamic symmetry, while the equivalence principle, previously considered to play a role only in gravity, plays a fundamental role in particle physics, in particular, it determines the strong CP-invariance.

\section*{Acknowledgments}
This research was supported by theme grants of Department of physics and astronomy of NAS of Ukraine: "Stochastic processes in condensed media, biological systems and radiation fields" 0125U000031 and by grant of Simons Foundation.


\begin{thebibliography}{99}

\bibitem{sad} Michael V. Sadovskii, \emph{Quantum Field Theory}, De Gruyter (2019), https://doi.org/10.1515/9783110648522

\bibitem{ryder} L.H. Ryder, \emph{Quantum Field Theory}, University of Kent, Canterbury (1987)

\bibitem{grig1} K.V. Grigorishin, Extended Time-Dependent Ginzburg–Landau Theory, J. Low Temp. Phys. \textbf{203} (2021) 262, https://doi.org/10.1007/s10909-021-02580-0

\bibitem{grig2} K.V. Grigorishin, Collective Excitations in Two-Band Superconductors, J. Low Temp. Phys. \textbf{206} (2022) 360, https://doi.org/10.1007/s10909-022-02668-1

\bibitem{grig3} K.V. Grigorishin, Collective excitations in three-band superconductors, Cond. Matt. Phys. \textbf{26} (2023) 23702, https://doi.org/10.5488/CMP.26.23702  

\bibitem{grig} K.V. Grigorishin, Three-band Extension for the Glashow–Weinberg–Salam Model, Acta Phys. Pol. B \textbf{56}, 8-A2 (2025),
https://doi.org/10.5506/APhysPolB.56.8-A2

\bibitem{abel} C. Abel, S. Afach et. al., Measurement of the permanent electric dipole moment of the neutron, Phys. Rev. Lett. \textbf{124}, 081803 (2020), https://doi.org/10.1103/PhysRevLett.124.081803

\bibitem{vilen} A. Vilenkin, E. P. S. Shellard, \emph{Cosmic Strings and Other Topological Defects}, Cambridge University Press, Cambridge (1994)

\bibitem{rubakov} V. Rubakov, \emph{Classical Theory of Gauge Fields}, Princeton University Press, Princeton (2002)

\bibitem{linde} A.D. Linde, \emph{Particle Physics and Inflationary Cosmology}, Harwood, Chur, Switzerland, 1990

\bibitem{dolgov}  A.D. Dolgov, Ya.B. Zeldovich, M.V. Sazhin, \emph{Cosmology of the early universe} (Moscow
Univ. Press, Moscow, 1988, in Russian).

\bibitem{kibble1} T.W.B. Kibble, Topology of cosmic domains and strings, J. Phys. A: Math. Gen. \textbf{9}, 1387 (1976), https://doi.org/10.1088/0305-4470/9/8/029

\bibitem{kibble2} M. B. Hindmarsh, T. W. B. Kibble, Cosmic strings,  Rep. Prog. Phys. \textbf{58}, 477-562 (1995), https://doi.org/10.1088/0034-4885/58/5/001

\bibitem{vilen1} A. Vilenkin, E.P.S. Shellard, \emph{Cosmic Strings and Other Topological Defects}, Cambridge University Press (1994)

\bibitem{vilen2} A. Vilenkin, Cosmic strings and domain walls, Phys. Rept. \textbf{121}, 263-315 (1985), https://doi.org/10.1016/0370-1573(85)90033-X

\bibitem{zurek1} W. H. Zurek, Cosmological experiments in condensed matter systems,  Phys. Rep. \textbf{276}, 177-221 (1996), https://doi.org/10.1016/S0370-1573(96)00009-9

\bibitem{zurek2} A. del Campo, W. H. Zurek, Universality of phase transition dynamics: Topological defects from symmetry breaking,  Int. J. Mod. Phys. A \textbf{29}, No.8 (2014), https://doi.org/10.1142/S0217751X1430018X

\bibitem{hind} M. Hindmarsh, A. Rajantie, Defect formation and local gauge invariance,  Phys. Rev. Lett. \textbf{85}, 4660-4663 (2000), https://doi.org/10.1103/PhysRevLett.85.4660

\bibitem{rajan} A. Rajantie, Formation of topological defects in gauge field theories,   Int. J. Mod. Phys. A  \textbf{17}, 1-44 (2002), https://doi.org/10.1142/S0217751X02005426

\bibitem{infirri} G. Sardo Infirri, D. Alesini, C. Braggio et. al., Search for Postinflationary QCD Axions with a Quantum-Limited Tunable
Microwave Receiver,  Phys. Rev. Lett. \textbf{135}, 211002 (2025), https://doi.org/10.1103/4dv9-72t5


\end{thebibliography}
\end{document}